\documentclass[sigconf]{acmart}

\usepackage{color}
\usepackage{soul}
\usepackage{graphicx}
\usepackage{amsmath}
\usepackage{xspace}
\usepackage{tikz}
\usepackage{enumitem}
\usepackage{threeparttable}
\usepackage{booktabs}
\usepackage{multirow}
\usepackage{ifthen}
\usepackage{cleveref}
\usepackage{subcaption}
\usepackage{balance}

\theoremstyle{definition}

\definecolor{color1}{rgb}{.2,.2,.2}
\newcommand*\circled[1]{\tikz[inner sep=.15ex,baseline=-.75ex] \node[circle,draw,color=white,fill=color1] {#1};}

\newcommand{\header}[1]{\par\vspace{-1mm}\medskip\noindent\textbf{#1.}}
\newcommand{\code}[1]{\texttt{\small #1}}
\newcommand{\smcode}[1]{\texttt{\scriptsize #1}}

\newboolean{showcomments}
\setboolean{showcomments}{true}
\ifthenelse{\boolean{showcomments}}
{
	\definecolor{myyellow}{RGB}{255, 228, 26}
	\definecolor{myblue}{RGB}{50, 50, 220}
	\newcommand{\nb}[2]{
		{\sf
			\fcolorbox{myyellow}{yellow}{\scriptsize\textbf{#1}}%
			$\blacktriangleright$%
			{\color{myblue}\fontsize{7pt}{8pt}\selectfont\textbf{#2}}%
		}%
	}
}
{
	\newcommand{\nb}[2]{}
}

\newcommand{\toolName}{\textsc{StyleX}\xspace}
\newcommand{\js}{\textsc{JavaScript}\xspace}
\newcommand{\html}{\textsc{HTML}\xspace}
\newcommand{\css}{\textsc{CSS}\xspace}
\newcommand{\crawljax}{\textsc{Crawljax}\xspace}

\newcommand{\totalNumberOfFeatures}{68\xspace}

\newcommand{\numberOfSubjectsRQTwo}{four\xspace}

\setcopyright{rightsretained}

\begin{document}
\title{
Style-Guided Web Application Exploration
}

\author{Davood Mazinanian}
\affiliation{%
	\institution{University of British Columbia}
	\city{Vancouver, BC}
	\state{Canada}
}
\author{Mohammad Bajammal}
\affiliation{%
  \institution{University of British Columbia}
  \city{Vancouver, BC}
  \state{Canada}
}
\author{Ali Mesbah}
\affiliation{%
	\institution{University of British Columbia}
	\city{Vancouver, BC}
	\state{Canada}
}



\begin{abstract}

A wide range of analysis and testing techniques
targeting modern web apps 
rely on the automated exploration of their state space
by firing events that mimic user interactions. However, finding out which elements are actionable in web apps is not a trivial task.
To improve the efficacy 
of exploring the event space of web apps,
we propose a browser-independent,
instrumentation-free 
approach based on structural and visual stylistic cues.
Our approach, implemented in a tool called \toolName, 
 employs machine learning models,  
trained on 700,000 web elements from 1,000 real-world websites, 
to predict actionable elements on a webpage a priori.
In addition, our approach uses stylistic cues
for ranking these actionable elements
while exploring the app.
Our actionable predictor models achieve
90.14\% precision and 87.76\% recall 
when considering the click event listener,
and on average, 
75.42\% precision and 77.76\% recall 
when considering
the five most-frequent event types.
Our evaluations show that
\toolName can improve
the \js code coverage achieved by a general-purpose crawler
by up to 23\%.

\end{abstract}

%
%

\maketitle


\section{Introduction}
\label{sec:introduction}

Modern web apps are built across different technology stacks, 
often using multiple programming languages, e.g., \js, \html,
Cascading Style Sheets (\css), and server-side languages such
as PHP or Java.
They are primarily written in an event-driven architecture,
in which event listeners on the webpage asynchronously
execute and change the state of the application in response to events
(e.g., clicks, mouse hovers)~\cite{Adamsen:2017}.

To mitigate the intricacies associated with analyzing 
web apps' sophisticated source code, 
a wide range of web app analysis and testing techniques
rely on \textit{automated exploration} of the
application in a black-box manner
~\cite{Mesbah:2012:InvariantBasedTesting, MilaniFard:2013:FeedEx,
MilaniFard:2014:LeveragingExistingTests, Mirshokrae:2015:JSFeet,
Duda:2008:AjaxSearchTool, Choudhary:2012:CrossCheck, Mesbah:2011:CrossT,
Thummalapenta:2013:GuidedTestGeneration, dallmeier2014webmate}.
In a nutshell, these techniques use a web crawler specifically designed 
 to mimic a user interacting with 
a web app's graphical user interface in a web browser. 
These crawlers automatically fire events on web elements,
fill in web forms, and monitor potential changes to the UI state 
to reverse-engineer an abstract model 
(e.g., a \textit{state-flow} graph~\cite{Mesbah:2012:Crawljax}) of the web app.
This inferred model is in turn used for various analysis and testing purposes, 
such as automated test case generation
~\cite{Thummalapenta:2013:GuidedTestGeneration, 
MilaniFard:2014:LeveragingExistingTests, Mirshokrae:2015:JSFeet}.

To achieve a sufficient degree of
\textit{state space coverage}~\cite{domcovery:issta14},
a crawler needs to effectively explore the event space. 
To that end, in each new state of the web app,
the crawler needs to
(1) identify web elements that can potentially trigger an event
(e.g., elements with click or mouseover events),
and (2) determine the order or sequence
of which web element to exercise next.

While the W3C event model standard~\cite{pixley2000document}
allows event listeners to be added to web elements
using the Document Object Model (DOM) API,
it does not provide access to the set of event listeners
already attached to a particular web element
~\cite{Andreasen:2017:SurveyOfDynamicAnalysis, Dallmeier:2013:WebMate}.
Hence, identifying web elements that have event listeners,
which we refer to as \emph{actionables} in this paper,
is not straightforward.

Crawlers vary in their strategies to circumvent this challenge.
For instance, they can be configured to click on all
elements that are known to be ``clickable'' by default,
such as hyperlinks (e.g., \code{<a href=``www.example.com''>})
 and buttons, which will then yield new states.
However, this can leave out a large number of actionables
which do not belong to these two element types,
and therefore can undermine the crawler's ability
 of discovering new states.
Indeed, research has shown that hyperlinks and buttons
are not the only doorways to new states in web
 apps~\cite{Behfarshad:2013:HiddenWeb},
and that other tags such as \code{<div>} and \code{<span>}
with event listeners are extensively used in practice.
Alternatively, configuring a crawler to explore
\emph{all} UI elements is also not a viable option,
since this will increase the
analysis and crawling time by examining
every element on each webpage
in search of actionables.
Most of this analysis might be wasted
since usually only a fraction of web elements are actionable.

Furthermore, a large set of existing crawlers
~\cite{Frey:2007-IndexingAjaxApps,
 Duda:2008:AjaxSearchTool, Moosavi:2013:ComponentBasedCrawling,
  Moosavi:2014:ComponentBasedCrawlingICWE}
only consider events that are attached using \html attributes 
(e.g., \code{onclick}, \code{onmouseover}) because such attributes
can be retrieved and accessed anytime, enabling the crawler
to identify which elements are actionable.
Other approaches (e.g., \textsc{WebMate}~\cite{Dallmeier:2013:WebMate, dallmeier2014webmate})
only consider actionables for which event listeners are
added using specific \js frameworks
(e.g, \textsc{jQuery} or \textsc{Prototype}).
However, event listeners are often attached to web elements
directly through the DOM API (e.g., \code{addEventListener()}).
Identifying actionables which use this technique
is more challenging as the DOM API
does not provide a mechanism for retrieving the registered handlers.
 
Common workarounds for identifying actionables are
 to use an instrumented browser engine 
(e.g., in \textsc{Artemis}~\cite{artzi2011framework}
 or \textsc{WaRR}~\cite{Andrica:2011:WaRR})
or specific browser add-ons (e.g., Chrome's 
DevTools~\cite{chrome-dev-tools}).
This, however, restricts the devised techniques
to a specific web browser, which is not desirable
in many scenarios (e.g., cross-browser 
testing~\cite{Mesbah:2011:CrossT, Choudhary:2012:CrossCheck}). 
This also limits such techniques to a specific browser engine's \textit{version}, which hinders keeping up 
with the rapid evolution of modern web browsers
to test new web platform features.

Another option is to instrument the web app's \js code, 
through a web proxy, and re-write the event listener add/remove calls.
These web proxies are known to be hard-to-deploy~\cite{Burg:2013:InteractiveRecordReply}.
Also, source code instrumentation
might undesirably alter the behaviour of the web app~\cite{Richards:2011:ConstructionOfJSBenchmarks}.
Hence, there is a need for a new approach
of identifying actionables for effective web app state exploration.

Once the set of actionables is identified,
the order in which the actionables are explored by the crawler
can also impact the state exploration of the web app.
Note that this is different from choosing the next web
app {state} to explore, from a given list of 
already-explored states (i.e., the state exploration strategy),
which has been studied in the
literature~\cite{MilaniFard:2013:FeedEx,
Benjamin:2011:StrategyHypercube:ICWE,
Choudhary:2012:MCrawler:Thesis, Choudhary:2013:MenuModel:ICWE,
Dincturk:2013:ModelBasedCrawlingStrategy,
Dincturk:2012:StatisticalStrategy,
Dincturk:2014:ModelBasedForRIA:TWEB,
Benjamin:2007:StrategyHypercube:Thesis}.
In contrast, the issue at hand here is concerned with
\emph{ranking} the execution order of
actionables on a given state.
Existing crawlers often fire events
in a random or top-down order~\cite{Mesbah:2012:Crawljax}.

In this paper, we show that these challenges
can be tackled in a rather novel way, namely by  
using the stylistic information of web elements.
In particular, we propose a technique that
identifies actionables based on the insight that 
a web element's \textit{structural and visual styles}
(i.e., their DOM location and the way they look)
can potentially indicate whether they have 
events attached to them.
In other words, we exploit the fact that 
a human user can easily, simply by ``looking'' at the webpage, 
spot which elements on the page can be clicked on.
Indeed, styling actionables in a consistent and 
distinguishing way is a highly-recommended usability practice~\cite{w3c-actionable-elements, bbc-actionables-usability-tips}.
We aim to capture this behavior in a model that can 
identify actionables based on their structural and visual styles.

We also show that these structural and visual stylistic
features can provide an effective
event ranking strategy during crawling to achieve a higher code coverage.
Our ranking approach essentially exploits
the Consistent Identification usability guideline~\cite{w3c-consistent-identification}:
elements with similar functionality should have a consistent presentation across the web app.
By postponing the execution of actionables that look similar,
we aim at diversifying the covered functionality.

To this end, we employ machine learning on a corpus of around
700,000 web elements collected from 1,000 websites 
to learn whether an element with specific stylistic features has an event listener.
We incorporate \totalNumberOfFeatures structural and visual style 
features for elements in the trained models.
Our approach can achieve 
90.14\% precision and 87.76\% recall in identifying clickable elements,
and on average, 
75.42\% precision and 77.76\% recall in identifying actionables 
of the five most-frequent event types.
Moreover, we devise an algorithm
for identifying similarly-looking elements
and ranking them accordingly to guide the crawler.
Our evaluations also show that our approach
can improve the covered client-side \js code
achieved by a crawler by up to 23\%.

Our paper makes the following contributions:

\begin{itemize}
	\item We provide a novel technique that uses structural and visual stylistic features of web elements
	to identify the ones with event listeners and rank them.
	\item A tool, \toolName, that builds upon a general-purpose web crawler for dynamic web apps
	to incorporate the proposed technique, which is available.
	\item A dataset of around 700,000 web elements 
	and their associated stylistic features as rendered in the web browser
	and the attached event listeners to them, which might be used in, e.g., future empirical studies.
\end{itemize}


\section{Background and Motivation}
\label{sec:motivation}

\subsection{The internals of modern crawlers}
\label{sec:background-crawlers}

\begin{figure}
	\centering
	\includegraphics[width=\linewidth]{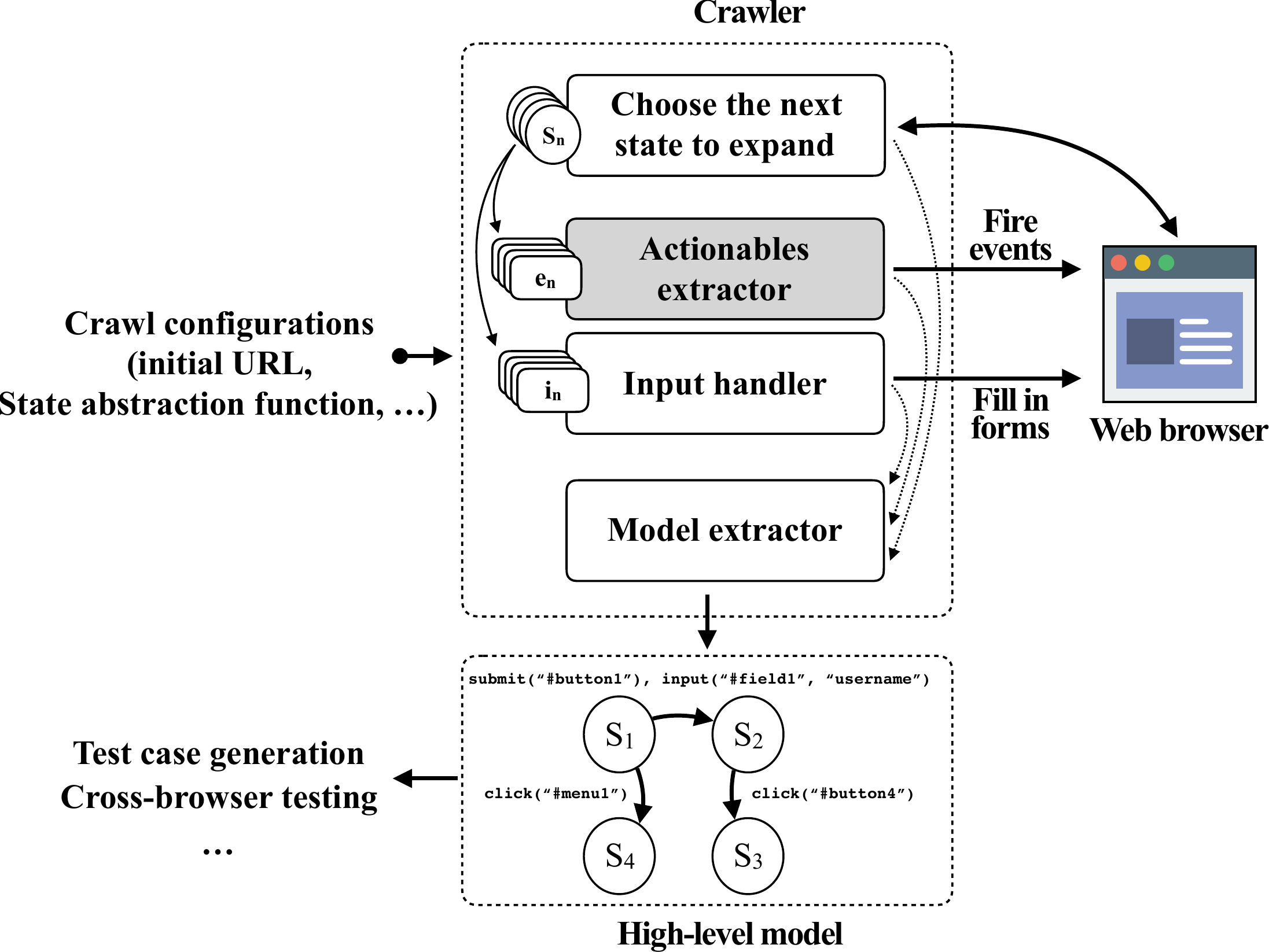}
	\caption{The internals of an abstract crawler.}
	\label{fig:abstract-crawler}
\end{figure}

\Cref{fig:abstract-crawler} illustrates the architecture of an abstract crawler
tailored for modern web apps.
The crawler starts by exploring the web app under analysis
from a starting point, e.g., the URL pointing to the web app's first page.
The crawler navigates a web browser to that URL,
and then extracts actionables in the loaded page.
These are DOM elements that the crawler could interact with
to change the \textit{state} of the application.
These elements are \textit{candidate} actionables
for a crawler, since it does not know whether interacting with them will 
lead to a new state a priori.
Subsequently, the crawler chooses the next event corresponding to a candidate actionable,
fires it,
and monitors the web browser to see whether there is a change in the current state of the web app.

Crawlers often allow custom \textit{state abstraction function}s; i.e.,
definitions for what constitutes a web app state
(e.g., a DOM snapshot, or a screenshot of the web page).
The crawler then chooses the next state to \textit{expand}
(i.e., to identify actionables on that state and continue the crawling from there)
based on a \textit{state exploration strategy}.
While crawling, the crawler can construct a model of the web app under analysis
using the states and transitions between them.
This can be a graph where each node is a state, 
and each edge is the event that caused the following state to be explored
(i.e., the \textit{state-flow graph}~\cite{Mesbah:2012:Crawljax}).
This model is then used for different purposes,
e.g., automated test case generation~\cite{Mirshokrae:2015:JSFeet}.

\begin{figure}[h]
	\centering
	\includegraphics[width=\linewidth]{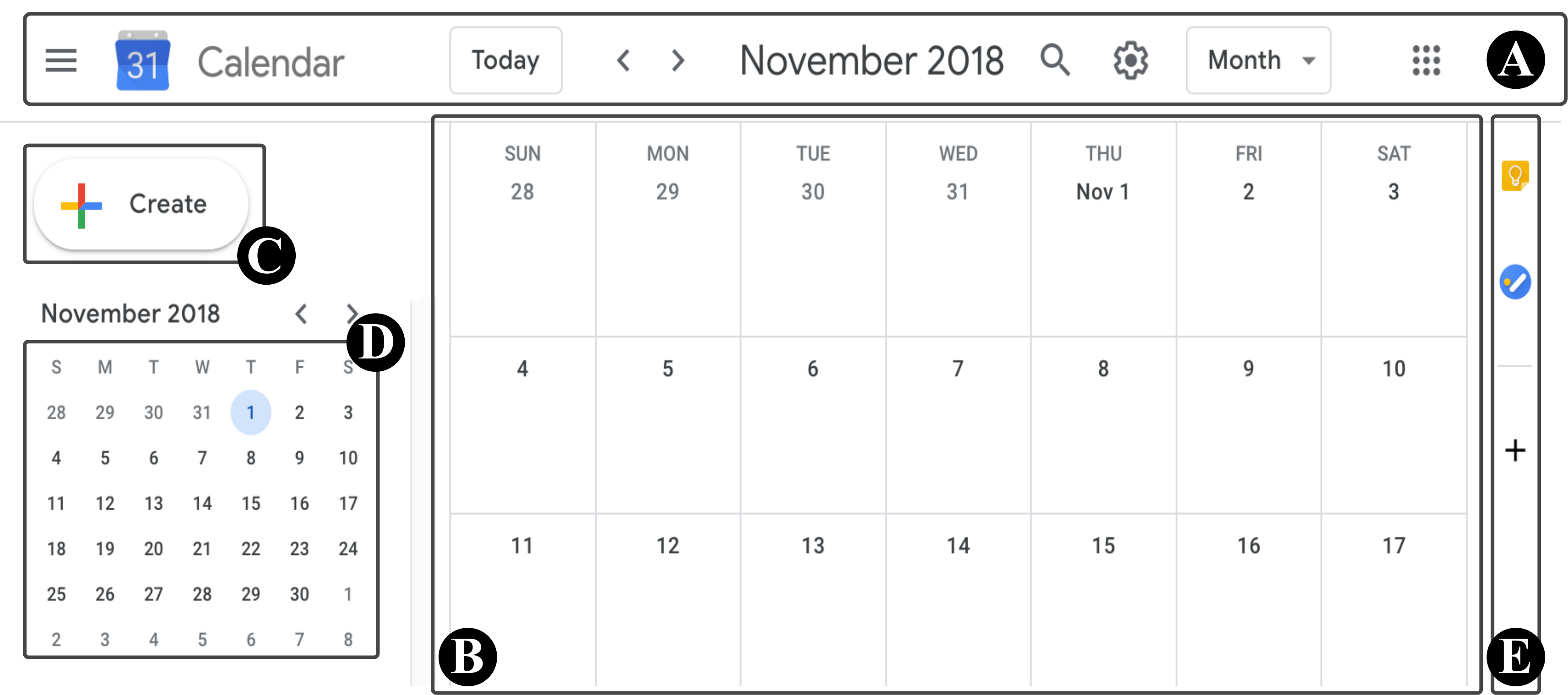}
	\caption{Interactive elements on Google Calendar's web UI.}
	\label{fig:motivating-example}
\end{figure}

\subsection{Motivating example}
\label{sec:motivating-example}

\Cref{fig:motivating-example} depicts a screenshot of Google Calendar, 
a highly-dynamic modern web app.
If we were to use an automated crawler to analyze Google Calendar
we would observe the following:


\header{Actionables}
	There is a large set of actionable elements that are candidates for interacting with
	to cover different functionalities.
	The \textit{actionables extractor} component of the crawler
	(i.e., the shaded box in \Cref{fig:abstract-crawler}) is responsible for identifying these elements.
	Most of these elements in Google Calendar are \code{<div>}s and not hyperlinks (i.e., \code{<a>} tags).
	This includes, for example, all elements in groups \circled{A}, \circled{B}, \circled{D} and  \circled{E} in \Cref{fig:motivating-example}. 
	Most web crawlers focus on hyperlinks and miss a large portion of actionables to be fired~\cite{Andreasen:2017:SurveyOfDynamicAnalysis}.
	More advanced crawlers~\cite{Mesbah:2012:Crawljax} give the user the option to specify the type of web elements (e.g., \code{div}) to consider as candidate actionables, which is still not ideal as the user has to manually configure the crawler.
	
	Our goal in this work is to devise a technique that predicts \textit{a priori} which elements on a page are actionables, 
	regardless of their element type and without manual configuration.
	
	\header{Equivalent classes of actionables} There are usually equivalent classes of actionables
	that appear within or across different states,
	and share similar functionality.  
	For instance, in Google Calendar, clicking on any of the displayed \textit{days} 
	(i.e., elements in group \circled{B} or \circled{D} in \Cref{fig:motivating-example})
	would result in seeing the calendar events corresponding to the selected day (elements in \circled{D}),
	or creating new calendar event in the selected day (elements in \circled{B}).
	In such cases where the corresponding functionalities are similar (or identical), 
	 firing events on these similar actionables is unlikely to result in new states
	or more \js code coverage.
	This can bear several implications in practice.
	If, for example, the crawler is configured to run within a certain time limit (which is the usual case in practice),
	it is highly probable that it would waste some of its runtime on those equivalent actionables. 
	In our motivating example of \Cref{fig:motivating-example}, 
	a more intelligent mechanism would be to 
	avoid firing events on the actionables in groups \circled{B} and \circled{D} again, 
	and moving on to another part of the web app,
	e.g., elements in groups \circled{A}, \circled{C}, or \circled{E} 
	which are likely to yield more different states and also cover a broader range of \js code.



\section{Approach}
\label{sec:approach}

Our approach for driving web app state exploration consists of two parts, namely, 
(1) a method for identifying actionables
using the web elements' structural and visual style features; and 
(2) a mechanism for ranking which events to fire
while exploring the states,  in order to to achieve a higher coverage.
In the following subsections,
we provide the details of these two steps.

\subsection{Predicting actionables}
\label{subsec:identifying-elements}
Our approach to identifying actionables
is based on a simple but novel intuition:
usually, when a user looks at a web page,
they are able to intuit where to click or
how to interact with the page based on what elements look like.
Styling actionables so that they look distinguishing
on the web page
is a recommended usability practice~\cite{w3c-actionable-elements, bbc-actionables-usability-tips}.
A clickable element, for example, might have an underlined text,
a border, or a shadow,
or might look like a button.
The intuition for our approach is that such stylistic
features can be used to train a machine learning model
to predict which elements on the page are actionable.

\begin{figure}
	\centering
	\includegraphics[width=\linewidth]{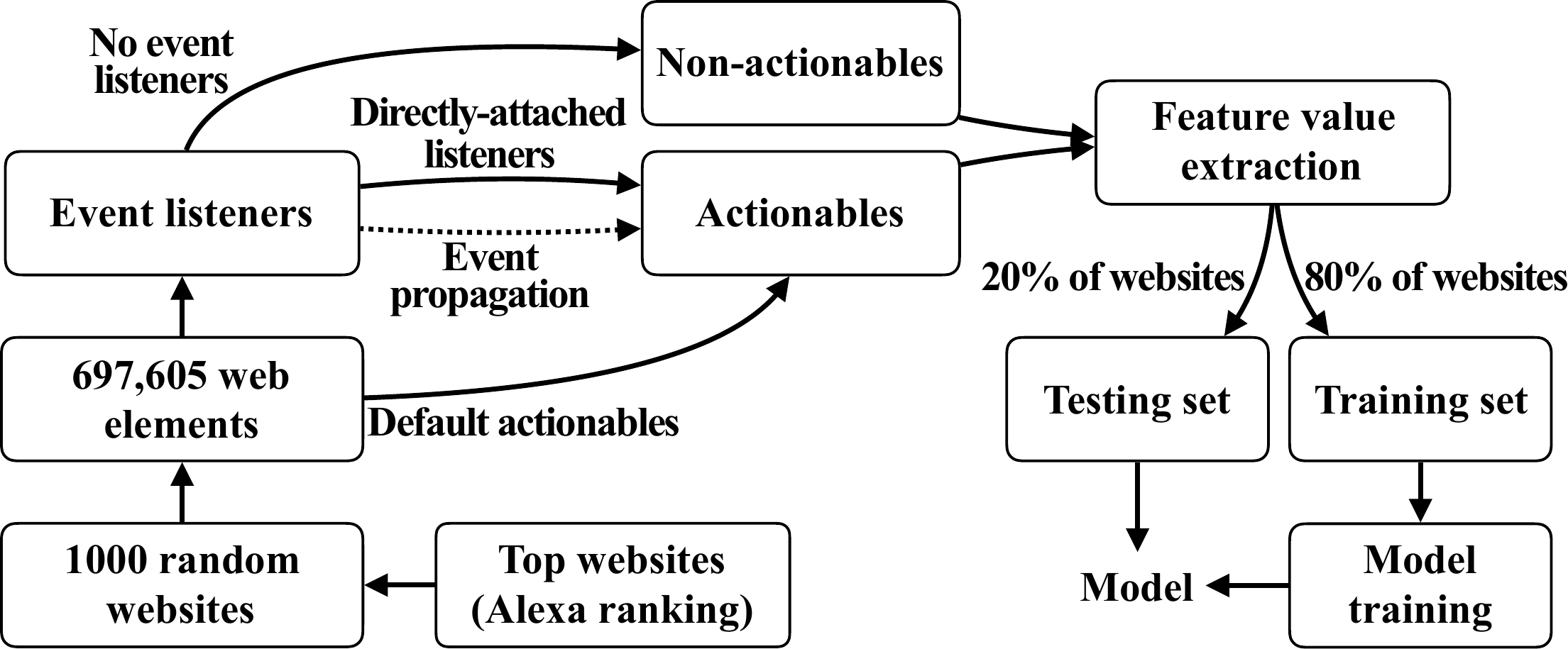}
	\caption{Our machine learning model training pipeline.}
	\label{fig:model-training-pipeline}
\end{figure}

\Cref{fig:model-training-pipeline} depicts our overall pipeline for learning.
We explain the details of our learning process in the following subsections.

\header{Data collection for training}
To collect data for training a model, 
we need a large set of pre-identified (1)  actionable elements to be used 
as positive examples, and (2) elements without any event listeners to be used as negative examples.

We collect these elements by crawling a large set of web apps in the wild.
To that end, we developed a script to randomly choose 1,000 websites
from Alexa's top ranking list of URLs~\cite{alexa-ranking}.
The reason for this random selection is 
to cover a wide range of sites 
that have been developed with different front-end frameworks, libraries, and development styles.

The script subsequently loads each URL in the Chrome web browser, 
traverses the DOM loaded in the web browser,
and collects all the \html elements present in the web page.
For each element, the script collects any attached event listeners.
The event listeners are retrieved using Chrome DevTools~\cite{chrome-dev-tools},
which have direct access to the browser's internal engine,
and thus is accurate. Note that this access is only needed for collecting the training data. 

If an element has an attached event listener,
it is considered \emph{actionable} and is stored for
later analysis along with the type of the event(s) handled.
In addition, we consider \html elements which are \textit{actionable by default}, 
such as hyperlinks and buttons, as positive examples, regardless if whether they have a \js event listener.
The reason for this is that these elements can change the state even without
an explicitly attached \js event.

\begin{figure}
	\centering
	\begin{subfigure}[b]{0.32\linewidth}
		\centering
		\includegraphics[scale=0.64]{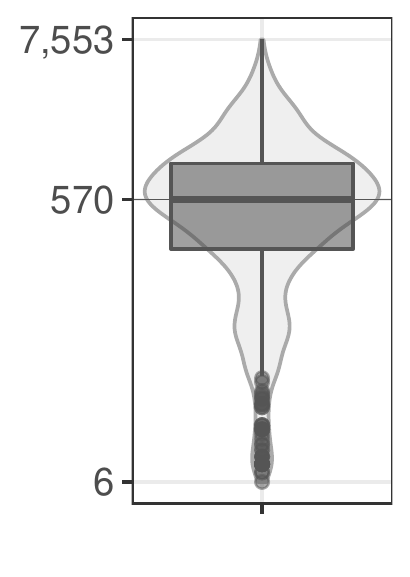}
		\caption{\#DOM nodes}
		\label{fig:number-of-dom-nodes}
	\end{subfigure}
	\begin{subfigure}[b]{0.32\linewidth}
		\centering
		\includegraphics[scale=0.64]{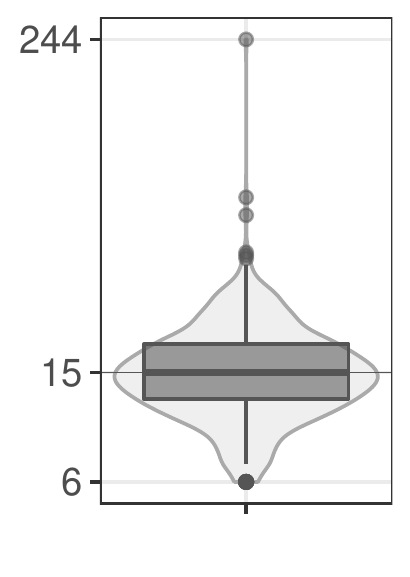}
		\caption{DOM height}
		\label{fig:dom-height-distribution}
	\end{subfigure}
	\begin{subfigure}[b]{0.32\linewidth}
		\centering
		\includegraphics[scale=0.64]{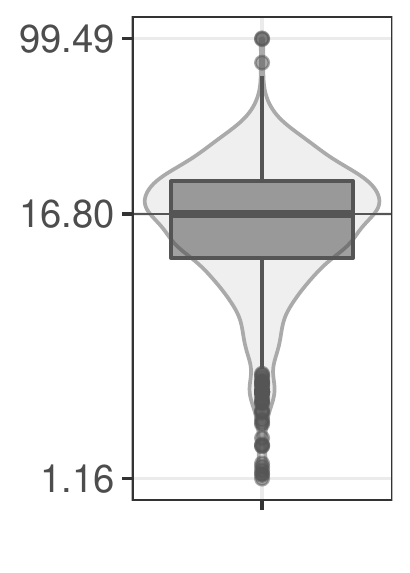}
		\caption{\%Actionables}
		\label{fig:percentage-elements-listeners}
	\end{subfigure}
	\caption{Descriptive statistics for the training websites.}
	\label{fig:websites-info}
\end{figure}

\Cref{fig:websites-info} depicts violin plots
summarizing the characteristics of the websites 
used in the learning process.
As it can be observed, the websites are quite complex
in terms of the number of DOM elements (\Cref{fig:number-of-dom-nodes})
and the height of the DOM tree (\Cref{fig:dom-height-distribution}).
In addition, we have shown the distribution of the percentage of the actionables
over all the DOM elements existing in each website
in \Cref{fig:percentage-elements-listeners}.
Notice that the median percentage of actionable elements is 16.80\%.
Moreover, we found that 53.26\% of these actionables 
are elements other than the \textit{default actionables} (i.e., hyperlinks and buttons).

\header{Incorporating event propagation}
Firing certain events on DOM elements 
causes the same event to be triggered on the element's ancestors.
For instance, when a user clicks on a button,
all the button's ancestor elements are also clicked too through event propagation.
The order in which the event listeners of the same type
attached to the ancestor elements
are executed can be different: 
i.e., first execute the events attached to the ancestors, going down in the DOM hierarchy---the \textit{capture} phase---%
or first execute the events attached to the element itself, going up in the hierarchy---the \textit{bubble} phase~\cite{event-bubble-capture}.

To incorporate this behavior in the training model,
for the event handlers that bubble,
we mark the descendant elements of an actionable
to be actionable too.
In our experience, this can greatly improve the accuracy of the trained models,
because the values for most of the structural and visual styles (i.e., the models' features)
are also \textit{inherited} by the descendant elements, through CSS inheritance~\cite{css-cascade-inheritance}.

\begin{table*}[h]
	\caption{Features used in the trained models}
	\centering
	\footnotesize
	\setlength\tabcolsep{3px}
	\begin{threeparttable}
		\bgroup
		\begin{tabular}{l p{15.5cm}}
			\toprule
			\textbf{Type} & \textbf{Features} \\ \midrule
			
			\textbf{DOM-related}	& 
				Bounding box position and size (x, y, width, height),
			    DOM depth, 
			    number of descendants,
			    height of the subtree rooted at the element \\ \midrule
			
			\textbf{Visual (\css)} &    
				\smcode{align-content}, 
				\smcode{align-items}, 
				\smcode{align-self}, 
				\smcode{backface-visibility}, 
				\smcode{border-block-end-style}, 
				\smcode{border-block-start-style}, 
				\smcode{border-bottom-style}, 
				\smcode{border-collapse}, 
				\smcode{border-inline-end-style}, 
				\smcode{border-inline-start-style}, 
				\smcode{border-left-style}, 
				\smcode{border-right-style}, 
				\smcode{border-top-style}, 
				\smcode{box-sizing}, 
				\smcode{clear}, 
				\smcode{cursor}, 
				\smcode{display}, 
				\smcode{flex-direction}, 
				\smcode{flex-grow}, 
				\smcode{flex-wrap}, 
				\smcode{float}, 
				\smcode{font-style}, 
				\smcode{font-weight}, 
				\smcode{hyphens}, 
				\smcode{justify-content}, 
				\smcode{list-style-position}, 
				\smcode{list-style-type}, 
				\smcode{mix-blend-mode}, 
				\smcode{object-fit}, 
				\smcode{opacity}, 
				\smcode{outline-style}, 
				\smcode{overflow-wrap}, 
				\smcode{overflow-x}, 
				\smcode{overflow-y}, 
				\smcode{pointer-events}, 
				\smcode{position}, 
				\smcode{resize}, 
				\smcode{table-layout}, 
				\smcode{text-align}, 
				\smcode{text-decoration-line}, 
				\smcode{text-decoration-style}, 
				\smcode{text-overflow}, 
				\smcode{text-rendering}, 
				\smcode{text-size-adjust}, 
				\smcode{text-transform}, 
				\smcode{transform-style}, 
				\smcode{unicode-bidi}, 
				\smcode{user-select}, 
				\smcode{visibility}, 
				\smcode{white-space}, 
				\smcode{word-break} \\
				
			& 
				In addition, binary values determining whether
				a value other than the default value 
				is set 
				for the following \css properties:
				Animation-related properties (i.e., \smcode{animation-name} and \smcode{transition-property}),
			    \smcode{background},
			    \smcode{border} and \smcode{outline} (at any of the four sides of the element),
			    \smcode{box-shadow},
			    \smcode{text-decoration},
			    \smcode{touch-action},
			    \smcode{transform},
			    \smcode{will-change}, and
			    \smcode{z-index}. \\
			\bottomrule
		\end{tabular}
		\egroup
	\end{threeparttable}
	\label{table:model-features}
\end{table*}

\header{Training features}
For each element,
we collect and store the values for a set of \totalNumberOfFeatures features
of structural and visual styles,
which will be used as learning features in the trained models.
\Cref{table:model-features} lists these features.
The values for these features are extracted via the DOM API.

There are two sets of such learning features:

\header{(Structural) DOM-related features}
These include:
\begin{itemize}[leftmargin=*]
	\item The absolute position (from the top-left corner of the web browser window)
	and size of the bounding box of the element
	as rendered in the web browser. 
	The rationale behind using the bounding box is that, 
	intuitively, a relatively large element is less probable to have certain event listeners.
	In addition, 
	an element with abnormal positioning (e.g., negative values, which corresponds to an element outside the viewport)
	is less likely to be actionable.
	We use the \code{Element.getBoundingClientRect()} DOM function to retrieve elements' bounding boxes.
	
	\item DOM depth, which corresponds to the depth of the element in the DOM hierarchy.
	The intuition here is that 
	the elements which are closer to the \html root (e.g., have smaller depth)
	are less likely to have event listeners to be actionable by users.
	Rather, they are mostly used in defining the structure of web pages.
	
	\item Number of descendants, and height of the subtree rooted at the element.
	These values determine the \textit{complexity} of the DOM subtree
	under the element.
	It is intuitive that elements which are more complex 
	(e.g., a \code{<nav>} element which is used as a container for menu items)
	are not actionable.
\end{itemize}

Note that, in our machine learning models,
we did not include elements' ``tag name'' as a feature.
The reason is that 
any element with any tag name can become actionable;
as a result, in practice we observed that including the tag name does not generally 
improve the trained models.
In addition, it is now possible to define \textit{custom \html elements}
with custom tag names,
both supported natively in all major web browsers in the 
Web Components standard~\cite{web-components},
and also in several popular frameworks (e.g., Angular~\cite{angular-custom-elements}).
In such cases, custom tag names would be unknown to the trained models.

\header{(Stylistic) \css features}
The latest \css specifications \cite{css-specs} include more than 200 \textit{style properties}.
Style properties are individual presentation features that can be set on any element
(e.g., font, border, background).
We use these stylistic properties 
as features for training models.

We retrieve the values for \css style properties 
using the DOM's \code{Window.getComputedStyle()} function.
This is a standard DOM function that is supported by all web browsers,
and frees us from dealing with complex \css code and its internals
(e.g., \css value propagation through inheritance and cascade~\cite{css-cascade-inheritance})
to compute the style property values.
In addition, the values returned by the \code{getComputedStyle()} function
are \textit{normalized} to a great extent
(e.g., all color values are represented using the \code{RGB} or \code{RGBA} notation),
making them suitable for machine learning models. 

We collected a list of 279 \css properties
as returned by the \code{getComputedStyle()} function from the elements of all websites.
We then excluded the \textit{vendor-specific} \css properties,
since they are recognized only in specific web browsers 
(e.g., properties prefixed with \code{--web-kit-} are only recognized by the WebKit engine, e.g., in Apple's Safari).
Our intention is to keep the trained models browser-agnostic.

For the remaining \css properties,
we plotted the distribution of their values across all websites.
We removed the \css properties for which the values for more than 99\% of elements were left as default.
This is because the values for these \css properties will not bring any benefit
for the machine learning models 
and therefore keeping them would unnecessarily complicate the training process.
We have provided the complete list of the remaining \css properties
which were eventually used in the training models in \Cref{table:model-features}.

Note that, for certain \css properties, 
it is not logical to use the raw values for trainig.
For example, the value for the \code{background-image} property
is set to a specific external image, or a gradient declaration,
and we observed that these values are not helpful for improving the models' accuracy.
Instead, it is more plausible 
to treat such values as binary predictors,
e.g., whether an element has a background or not.
In addition to the \code{background-image} property,
the following \css properties are treated as binary predictors:
\code{animation},
\code{border},
\code{box-shadow},
\code{outline},
\code{text-decoration},
\code{touch-action},
\code{transform},
\code{wi\-ll-change}, and
\code{z-index}.


\header{Training and testing sets} We divide the set of collected websites
into two subsets for training and testing,
with an 80\%-20\% split, respectively, so that it allows for cross validation,
and use the corresponding positive and negative examples for training and testing.

\header{Balancing positive/negative examples} There are usually fewer positive examples 
than negative ones in each web page.
This can affect the accuracy of the trained models.
Therefore, we balanced the number of positive examples 
with the negative ones
using under-sampling, i.e., randomly removing elements from negative examples
until the two sets have the same number of elements.

\header{Choosing event types}
There is a large number of event types supported in the web platform standards
(e.g., click, mouse, keyboard, touch, drag, and change events).
In addition, developers can define \textit{custom} event listeners.
In this work, we focus on the five most frequent event types
that appeared in our dataset, 
namely \code{click}, \code{mouseover}, \code{mouseout}, and \code{mousedown}, and \code{touchstart}.
For each of these event types, we train a separate binary model
that predicts whether a certain element has a listener for this given event type.

\header{Machine learning algorithms}
In our experiments, we used several different traditional machine learning algorithms  
for classification
which met our input and output requirements.
This includes, but is not limited to, CART, C4.5, and C5.0 decision trees,
random forests,
and feed forward neural networks.
For our experiments, we eventually selected the model with the highest accuracy
and deployed it to a general-purpose crawler.

\subsection{Prioritizing actionables using styles}
\label{sec:prioritization}

As discussed in \Cref{sec:motivating-example},
the appearance of actionables
can be used as a cue
towards improving the effectiveness of web app state exploration, for instance with respect to coverage.
Our intuition is that actionables with dissimilar appearance 
might be better candidates to be examined earlier, 
since they might represent the same functionality in the web app.
This intuition indeed stems from the Consistent Identification usability guideline~\cite{w3c-consistent-identification},
which aims at ensuring ``consistent identification of functional components
that appear repeatedly within a set of Web pages''.
As such, the exploration should ideally avoid exercising similarly-looking elements 
within the same state or across different states.

\header{Representing actionables}
The goal of our prioritization approach is to identify elements with similar appearance 
across different states of the web app,
so that when the crawler comes across a new actionable,
it can be ranked with respect to the 
ones that have already been explored.
To do so, we represent actionables in such a way that 
they can be compared across different states.

For purposes of ranking, each actionable
is represented using a vector of features $\hat{f_E}$
consisting of stylistic properties.
The elements of $\hat{f_E}$ are similar to the ones used
for training actionable prediction models
(\Cref{table:model-features}).
In particular, $\hat{f_E}$ contains all the \css properties that we used for prediction,
but excludes most of the DOM-related properties
(i.e., the position of the actionable's bounding box, 
depth of the actionable in the DOM tree,
number of descendants and height of the DOM subtree rooted at the actionable).
The reason we exclude DOM-related properties is that
the values of these properties for the same actionable can frequently change across different DOM states,
e.g., when an element is dynamically injected into the DOM 
at runtime using \js.
We experienced that including these properties
can negatively affect the effectiveness of this representation
in identifying the same (or similarly-looking) actionables across different states.
In contrast, 
we noticed that including concrete values for the \css properties
that we treated as binary predictors (i.e., last row of \Cref{table:model-features})
can improve this representation.
As such, we further enriched $\hat{f_E}$ with those \css properties.

\header{Ranking actionables}
We propose to rank actionables identified by the 
crawler's \textit{actionables extractor} (\Cref{fig:abstract-crawler}) 
in such a way that more diversely-looking elements are examined first.
In our approach,
the crawler maintains a global list $\mathbb{L}$ of tuples $\langle \hat{f_E}, c_E \rangle$, 
where $\hat{f_E}$ is the stylistic representation for a set of similarly-looking actionables $E$
(as described earlier),
and $c_E$ is a counter.
The elements of $\mathbb{L}$ indeed correspond to the already-examined actionables during a crawling session,
where $c_E$ corresponds to the number of times any actionable $e \in E$ represented by $\hat{f_E}$
has been examined by the crawler in the current crawling session.

Each time an actionable $e$ is examined by the crawler,
its corresponding feature vector $\hat{f_{e}}$ is constructed
and compared against the feature vectors existing in $\mathbb{L}$.
If there exists a $\langle \hat{f_E}, c_E \rangle \in \mathbb{L}$ where 
$\delta(\hat{f_{e}}, \hat{f_E}) < \epsilon$,
$c_E$ is incremented,
or else $\langle \hat{f_{e}}, 1 \rangle$ is added to $\mathbb{L}$.
Here, $\delta$ is a distance function 
and $\epsilon$ is a threshold,
which determine the degree of stylistic dissimilarity that is allowed 
for $e$ to be from the rest of actionables in $E$.

Similarly, in each new state,
the crawler queries $\mathbb{L}$ to rank a newly-identified actionable $e^{\prime}$:
if the feature vector corresponding to $e^{\prime}$ is similar enough
to a feature vector existing in $\mathbb{L}$,
say $\langle \hat{f_E}, c_E \rangle$,
it is ``pushed back'' for later examination.
In this case, the degree to which the examination of $e^{\prime}$ is delayed will depend on the value of $c_E$;
the lower $c_E$ is, the earlier $e^\prime$ will be examined by the crawler.


\section{Implementation}
\label{sec:implementation}

We automated our data collection using Chrome DevTool protocol's
open-source implementation in Java (\textsc{cdp4j}~\cite{chrome-dev-tools-java}).
We trained the machine learning models
using \textsc{R}~\cite{r}.
We used the implementation of C5.0 model in R~\cite{c50}
as the final classifier,
as it showed the best results in our evaluation.

We developed \toolName,
a tool that extends a general-purpose crawler, namely \crawljax~\cite{Mesbah:2012:Crawljax}, 
which is tailored for exploring highly dynamic web apps. \toolName enhances \crawljax's \textit{candidate actionables strategy}
by incorporating the trained models
to better identify actionables,
and also by ranking the actionables 
using stylistic cues
to improve state exploration.

To deploy the trained models from R to Java,
we used rJava~\cite{rjava}, which allows communicating with the R environment from Java.
All the developed tools, 
collected data for training,
evaluation results, 
and the training/testing data are available online~\cite{experimental-data}.


\section{Evaluation}
\label{sec:evaluation}

To evaluate our proposed technique,
we design a study aiming at answering
the following research questions:

\begin{enumerate}[label=\textbf{RQ\arabic*}, leftmargin=*]
	\item What is the accuracy of \toolName in identifying actionables?
	\item Does \toolName improve the
	exploration of web apps in terms of coverage?
\end{enumerate}

\subsection{RQ1: Accuracy of the actionable prediction models}

\subsubsection{Study design}
As mentioned,
we randomly choose 20\% of the websites 
from our initial dataset
for our testing set,
which includes 139,521 elements,
out of which 74,129 are actionable
(incorporating the event propagation).
We report the accuracy of the trained models
using the following measures commonly used in
evaluating the performance of machine learning models:

\header{Precision} The \textit{precision} of a classification model
for a class $c$
is defined as the ratio of the instances that the model correctly classifies 
to belong to the class $c$
(i.e., true positives)
compared to all the instances that the model classifies (correctly and incorrectly) to belong to $c$.
The precision is therefore calculated as
$\text{Precision}_{c} = \frac{TP_{c}}{TP_{c} + FP_{c}}$,
where $FP_{c}$ (i.e., false positives) are the cases that have been incorrectly classified by the model
to belong to class $c$.

\header{Recall} The \textit{Recall} of the classification model
for a class $c$
is defined as the ratio of the cases that the model correctly classifies 
to belong to class $c$
over all the cases that actually belong to class $c$.
In other words, $\text{Recall}_{c} = \frac{TP_{c}}{TP_{c} + FN_{c}}$,
where $FN_{c}$ (i.e., false negatives) are the cases 
where the model have missed classifying them to class $c$.

\header{F-measure} The harmonic mean of precision and recall for a class $c$, 
defined as $\text{F-measure}_c = 2 \times \frac{\text{Precision}_c \times \text{Recall}_c}{\text{Precision}_c + \text{Recall}_c}$.

\subsubsection{Results}
Our experiments showed that the C5.0 decision tree
with 10 boosting iterations~\cite{kuhn2013applied}
is by far the most accurate model out of the various models we
experimented with: CART, C4.5, C5.0 decision trees, random forests,
and feed forward neural networks.
Due to space limitations, we only report the results for the models 
trained using C5.0.
The data and the scripts for replicating the training and testing
for other models is available~\cite{experimental-data}.
Also, since we are training binary classifiers,
we report the results with respect to two classes:
actionables, and non-actionables.

\begin{table}
	\caption{Accuracy of the C5.0 actionable prediction model}
	\centering
	\footnotesize
	\setlength\tabcolsep{3px}
	\begin{threeparttable}
		\bgroup
		\begin{tabular}{c r r r c r r r}
			\toprule
			& 
			\multicolumn{3}{c}{\textbf{Actionable}} & &
			\multicolumn{3}{c}{\textbf{Non-actionable}} \\
			\textbf{Event Type} & 
			{Precision} & {Recall} & {F-measure} & &
			{Precision} & {Recall} & {F-measure} \\ \midrule
			
			\textbf{\smcode{click}} &
			90.14 & 87.76 & 88.93 & &
			87.91 & 90.26 & 89.07 \\
			
			\textbf{\smcode{mouseover}} &
			75.45 & 76.80 & 76.12  & &
			92.85 & 92.34 & 92.59  \\

			\textbf{\smcode{mouseout}} &
			74.79 & 72.13 & 73.44  & &
			91.39 & 92.41 & 91.90  \\
			
			\textbf{\smcode{mousedown}} &
			72.54 & 78.00 & 75.17  & &
			96.55 & 95.43 & 95.99  \\
			
			\textbf{\smcode{touchstart}} &
			64.18 & 74.12 & 68.79  & &
			96.67 & 94.90 & 95.78  \\ \midrule
			
			\textbf{Average} &
			75.42 & 77.76 & 76.49 & &
			93.07 & 93.07 & 93.06 \\

			 \bottomrule
		\end{tabular}
		\egroup
	\end{threeparttable}
	\label{table:rq1-accuracy}
\end{table}

\Cref{table:rq1-accuracy} shows the accuracy measures for the C5.0 model
when tested on the websites in the testing set.
As it can be observed, the trained model can achieve a remarkably high precision and recall
(respectively, 90.14\% and 87.76\%)
for predicting whether an element has an attached \code{click} event listener or not.
The precision and recall drops for other event types for the ``actionables'' class,
respectively,
75.45\% and 76.80\% for \code{mouseover},
74.79\% and 72.13\% for \code{ouseout},
72.54\% and 78\% for \code{mousedown},
and finally,
64.18\% and 74.12\% for \code{touchstart}.
This is rather expected since there are fewer true examples in the dataset
for these event types, 
and therefore the training is done with less available data.
Overall, our prediction model achieves an average of
75.42\% precision and
77.76\% recall (76.49\% overall F-measure)
when predicting all five event types. 

\header{Importance of the predictors}
We further analyzed the prediction models 
to find out which features
are the most important 
in predicting whether an element is actionable or not.

For C5.0 decision tree,
a common approach to determine the importance of the predictors 
is to consider the percentage of training set samples that fall into all the decision trees' terminal nodes after the split.
Using this measure we noticed that, for all the five event types,
three predictors play an important role in prediction:
DOM depth,
\code{cursor},
and the position of the bounding box of the element.

The importance of the depth of an element in the DOM tree can be explained by the fact that,
often,
the elements which are closer to the leaf nodes of the DOM tree
are more probable to be actionable.
Instead, elements such as \code{<html>} or \code{<body>}
with very small depth values
are less probable to be actionables,
when considering a large set of web apps in the wild.

Similarly, the importance of the \code{cursor} predictor is quite expected.
Especially for mouse events (which constitute four of the five event types that we considered in this paper),
the shape of the mouse cursor when an element is hovered on can effectively give a hint to the users
that the element is actionable and a mouse event is probably handled for it
(e.g., a cross-shaped cursor with arrows
might correspond to a \code{mousedown} event to start moving an element).

The position of an element can also be a strong predictor
for actionables.
Elements with out-of-the-viewport positioning (e.g., negative coordinates)
cannot be interacted by the user.
We noticed that such elements exist in the training web apps when,
for instance,
the web app uses an image carousel in its design.
Also, elements attached to the top-left corner of the web browsers' viewport
are less probable to be actionable.
This includes, for instance, the \code{<html>} and \code{<body>} elements,
or a navigation bar:
while the elements inside the navigation bar are usually actionable, 
the top-level container of the navigation bar is less probable to be actionable.

\subsection{RQ2: Coverage efficiency of the proposed technique}

\subsubsection{Study design}
\label{sec:rq2-study-design}

We evaluate the efficiency of the proposed technique 
in terms of the improvement in achieving higher \js code coverage
as discussed in the following subsections.

\header{Experimental subjects}
We included \numberOfSubjectsRQTwo subject applications for our evaluation.
Note that none of these evaluation subjects 
were used 
for training our actionable prediction models.
In addition to three open source \js web apps,
we have included one real-world,
extensively-used,
highly-dynamic proprietary web app
for our evaluation to assess the effectiveness of the proposed technique
for scenarios in which using an in-house server or execution environment
is not possible or desirable,
e.g., when functional testing of the web app is outsourced,
which is a common practice in the industry.
\Cref{table:experimental-subjects} shows the characteristics of our experimental subjects.

\begin{table}[b]
	\caption{Experimental subjects}
	\centering
	\footnotesize
	\setlength\tabcolsep{3.5px}
	\begin{threeparttable}
		\bgroup
		\def\arraystretch{1.2}
		\begin{tabular}{c l  r  r r  r}
			\toprule
			\multicolumn{3}{c}{} &
			\multicolumn{2}{c}{\textbf{\%Actionable nodes}\tnote{\textdaggerdbl}} &
			\\ \cmidrule{4-5}
			
			&
			\textbf{Name} &
			\textbf{\#DOM nodes}\tnote{\textdagger} &
			\textbf{All} & \textbf{Default} &
			\textbf{JS (KB)}\tnote{$\ast$}   \\ \midrule
			
			\multirow{4}{*}{\rotatebox[origin=c]{90}{Open src.}}
			& \textbf{p4wn (Chess)}     		& 379.00            & 69.12 & 18.46 & 7.12           \\
			& \textbf{TacirFormBuilder} 		& 175.69            & 46.31 & 10.56 & 338.61             \\ 
			& \textbf{Phormer Photo Gallery} 	& 292.51            & 37.71 & 16.91 & 52.37              \\[4pt]
				
			\multirow{2}{*}{\rotatebox[origin=c]{90}{Prop.}}
			& \multirow{2}{*}{\textbf{Google Calendar}}  		& \multirow{2}{*}{1622.34}          & \multirow{2}{*}{28.56} & \multirow{2}{*}{3.99} & \multirow{2}{*}{8,761.71}            \\ 
			& & & & & \\ \bottomrule
		\end{tabular}
		\egroup
		\begin{tablenotes}
			\item[\textdagger] Average number of DOM nodes across all states discovered during 5 executions of the crawler with all configurations.
			\item[\textdaggerdbl] Average percentage of DOM nodes which are actionable across all crawling sessions.
			Default actionables include anchors and buttons.
			\item[$\ast$] Size of the maximal set of \js code blocks discovered across all crawling sessions.
		\end{tablenotes}
	\end{threeparttable}
	\label{table:experimental-subjects}
\end{table}

\header{Study setup}
We run a general-purpose crawler (namely, \crawljax~\cite{Mesbah:2012:Crawljax})
on all subject systems,
using different strategies
for identifying and ranking actionables.
We have built our technique on top of \crawljax
so that the confounding factors that might affect the efficacy of our technique and its rival strategies 
are controlled to the largest possible extent. 
Furthermore, since \crawljax only uses click events by default, 
for a fair comparison,
we first run our technique only using click events,
but we also report the results with other event types separately.

Particularly, we run \crawljax with the following strategies:

\begin{enumerate}[leftmargin=*]

\item \textbf{Default clickables (DEF)}.
In this strategy, the crawler only clicks on elements
which are \textit{clickable} by default. 
This includes hyperlinks (i.e., all the \code{<a>} tags)
and buttons (including all \code{<button>} tags, 
and \code{<input>} tags with their \code{type} property
set to one of the following values: \code{button}, \code{submit}, or \code{image}).
These elements are clicked on in the order 
they appear in a preorder traversal of the DOM tree
(i.e., the output of the DOM's \code{getElementsByTagName()} function).

\item \textbf{All elements, random order (RND)}.
In this strategy the crawler clicks
on all the \html elements (in a given state)
in a random order.

\item \textbf{\toolName, click-only events (\toolName-CLK)}.
We run our proposed \toolName approach (i.e., \crawljax enhanced by our technique for predicting actionables
and ranking them),
yet we only allow clicking on elements
for a fair comparison with the previous strategies.
In this paper, we evaluate our technique using strict similarly of style feature vectors,
i.e., the $\delta$ function defined in \Cref{sec:prioritization}
checks whether two feature vectors are element-wise identical ($\epsilon = 0$) or not.

\item \textbf{\toolName, with all five event types (\toolName-EVNTS)}
Similar to {\toolName-ORD}, 
yet we allow all five event types predicted on the elements
to be examined by the crawler.
For events that require input
(e.g., \code{mousedown} which can determine which mouse button was clicked),
we provide random values.
In case if an element is predicted to be actionable
with multiple event types,
we rank the events with respect to their popularity in our dataset
(i.e., in the following order: \code{click}, \code{mouseoever}, \code{mouseout}, \code{mousedown}, and \code{touchstart}).

\end{enumerate}

For all the subjects, we start the crawler from the first page of the application.
For Google Calendar, we start the crawler on the first page appearing right after the user has logged in,
so that the full functionality of the web app is available to the crawler
(we created a fresh user account for this evaluation).
We do not limit the maximum crawling depth
or the maximum number of states discovered by the crawlers.
We do not change the \textit{state abstraction function} of \crawljax
across all the evaluations;
a new UI state is discovered whenever there is a change in the DOM~\cite{Mesbah:2012:Crawljax}.
In addition, we run the crawler with each of the strategies five times on each subject,
and report \js code coverage as the average across the five runs.
We do not define any strategy to provide inputs for web forms
during the crawl.
Furthermore, we run the crawlers in two scenarios:

\begin{enumerate}[leftmargin=*]

\item We limit the crawler's running time, for each of the four mentioned strategies (i.e., DEF, RND, \toolName-CLK, \toolName-EVNTS),
to 10 minutes. 
This duration should be relatively reasonably long enough to explore crawling behaviour,
but is manageable enough to enable repeating the experiment multiple times to compute averages.
Relevant work~\cite{MilaniFard:2013:FeedEx} also use this time limit for running crawling experiments. 

\item We also allow the crawler with each of the four mentioned strategies
to execute 100 \emph{crawl actions} 
(i.e. firing events on the states),
and measure \js code coverage right after each crawl action.
The rationale for this evaluation approach is that
we would like to compare the ability of crawling techniques 
in making \textit{good incremental decisions}
while crawling, 
i.e., examining the most efficient actionable firing order.
An efficient firing order would quickly yield higher coverage,
in a small number of firings, and does so 
early on in the crawling process.
In addition,
if the goal of the crawling is to generate test cases,
the number of the generated test cases will correlate with the number of crawl actions,
and our evaluation can reveal
how one strategy performs compared to the others
given a similar number of generated test cases,
as explored in an existing relevant work~\cite{artzi2011framework}.

\end{enumerate}

\begin{figure*}
	\centering
	\begin{subfigure}[b]{\linewidth}
		\centering
		\includegraphics[width=\linewidth, trim={0 1.25cm 0 0}, clip]{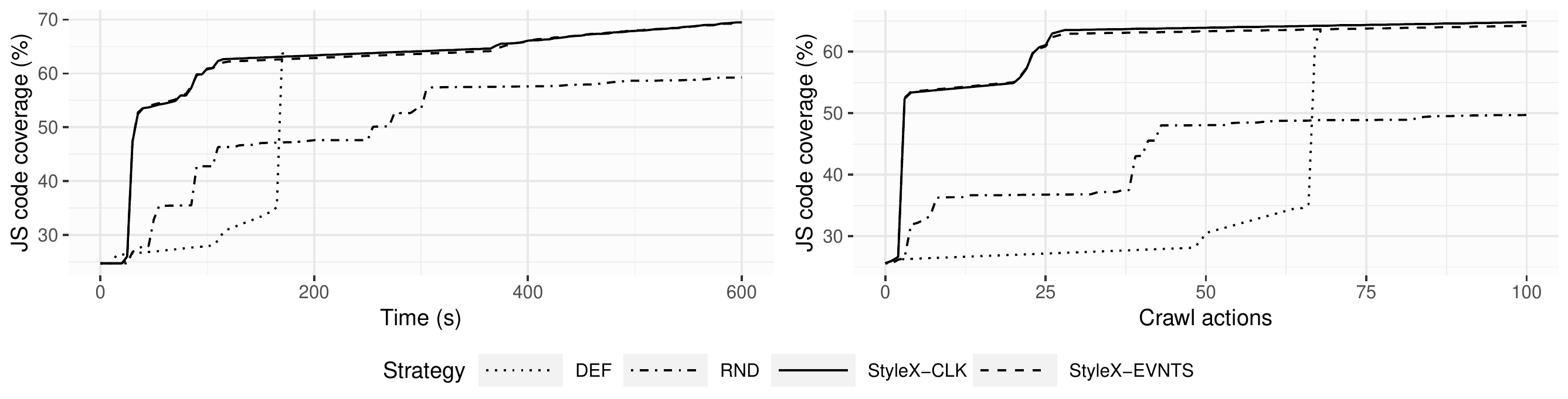}
		\vspace{-6mm}
		\caption{p4wn (Chess)}
		\label{fig:coverage-chess}
	\end{subfigure}\\[8pt]
	\begin{subfigure}[b]{\linewidth}
		\centering
		\includegraphics[width=\linewidth, trim={0 1.25cm 0 0}, clip]{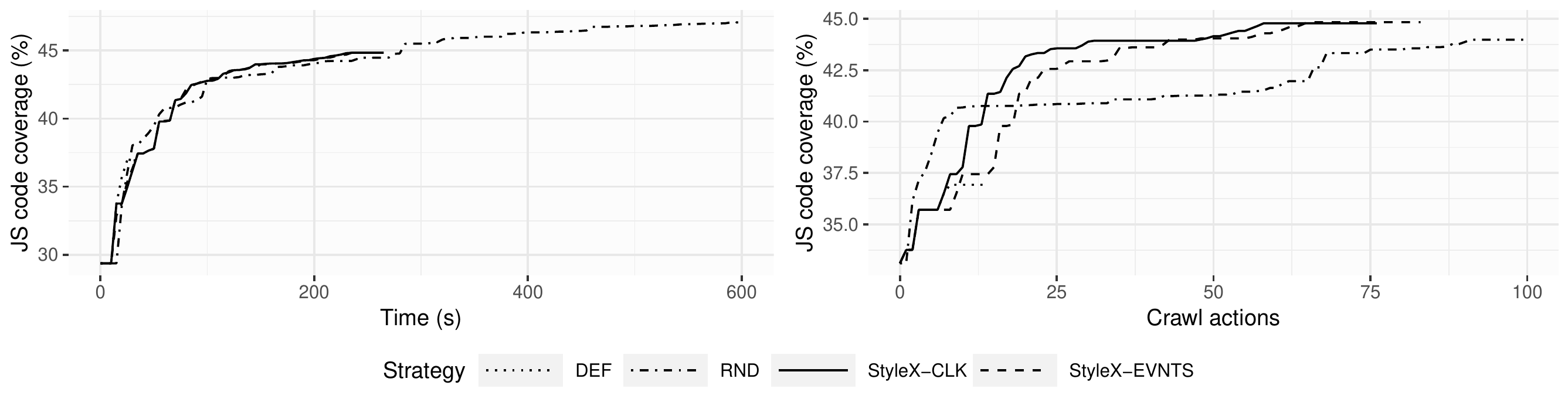}
		\vspace{-6mm}
		\caption{TacirFormBuilder}
		\label{fig:coverage-formbuilder}
	\end{subfigure}\\[8pt]
	\begin{subfigure}[b]{\linewidth}
		\centering
		\includegraphics[width=\linewidth, trim={0 1.25cm 0 0}, clip]{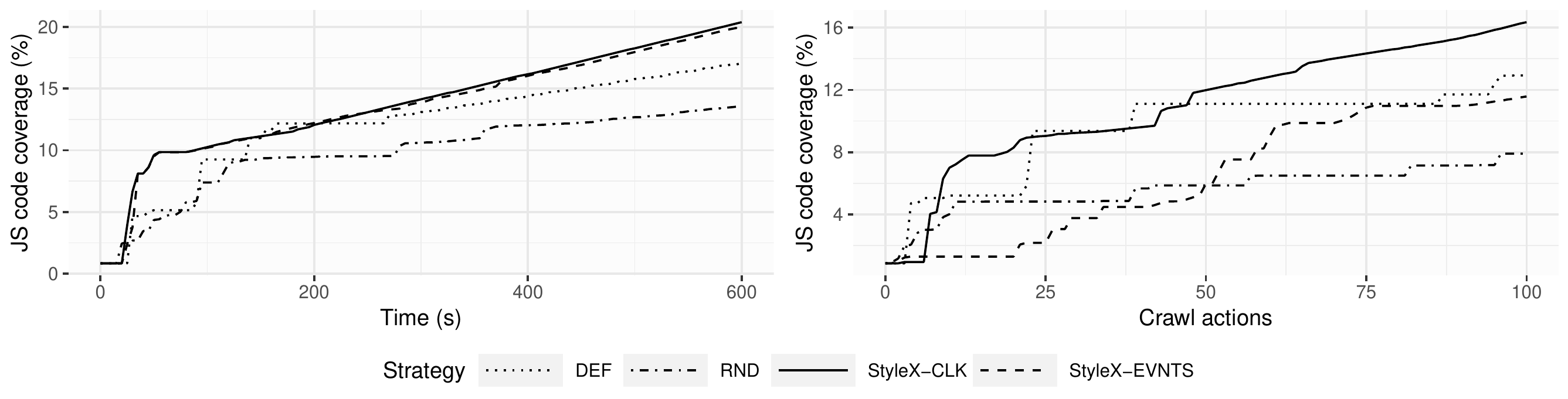}
		\vspace{-6mm}
		\caption{Phormer Photo Gallery}
		\label{fig:coverage-phormer}
	\end{subfigure}\\[8pt]
	\begin{subfigure}[b]{\linewidth}
		\centering
		\includegraphics[width=\linewidth]{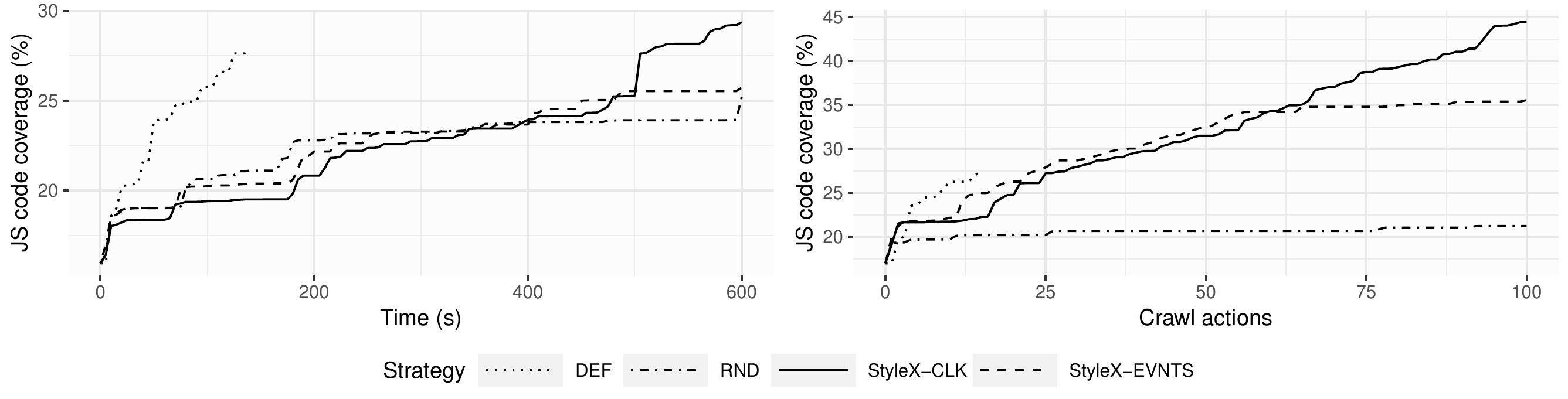}
		\vspace{-4mm}
		\caption{Google Calendar}
		\label{fig:coverage-calendar}
	\end{subfigure}
	\caption{Coverage performance.}
	\label{fig:coverage-results}
\end{figure*}

\header{Measuring \js code coverage}
Measuring \js code coverage for dynamic web apps
is not a straightforward task.
For a crawler, the total number of lines of \js code
is typically unknown,
as \js code can be
injected dynamically at runtime
while exploring new states.
Previous work have disregarded the dynamically-injected code blocks 
when measuring code coverage~\cite{artzi2011framework};
however, we argue that they should be considered because they do occur
in reality,
especially for highly-dynamic web apps.
Therefore, 
we consider the largest set of \js code lines
that is found by any strategy during the crawl
(after all the crawl sessions are finished)
as the maximal set of lines of code that should be covered by the crawler
(i.e., a lower bound for the actual size of the \js codebase).
We then compute the code coverage as the percentage of code 
that is covered by each strategy over the size of the \js code in this maximal set.

We use the precise code coverage tool provided by
Chrome DevTools~\cite{chrome-dev-tools}.
Note that, since \js code can be obfuscated (i.e., shortened, concatenated into a single line), 
we report the coverage with respect to the number of characters covered,
as opposed to lines of code.
The \js size reported in \Cref{table:experimental-subjects} is calculated from this value.
This also includes the embedded and inline \js code blocks
(i.e., the code enclosed in the \code{<script>} tags and in the \html attributes concerning event listeners, e.g., \code{onmouseover}).

\subsubsection{Results}
We have illustrated the code coverage achieved
using different crawling strategies
in \Cref{fig:coverage-results},
for the constraints on time (left) and the number of crawl actions (right).

Observe in \Cref{fig:coverage-results} that, for all subjects,
the crawler finishes before the maximum time/number of crawl actions is reached
when only default clickables are considered (i.e., the \textit{DEF} strategy).
This essentially stresses the importance of considering elements 
other than default clickables when crawling,
also supporting the conclusions of an existing work~\cite{Behfarshad:2013:HiddenWeb}
on the importance of other element types for covering web app's state space.
Notice how including only clickables has led to achieving less final \js code coverage
compared to using our proposed technique.
Notwithstanding, we can see that in Google Calendar,
using default clickables yields more coverage earlier in crawling,
in contrast to other subjects.
This suggests that giving more priority to default clickables will not necessarily 
improve the crawling in terms of \js code coverage.

Also, observe in \Cref{fig:coverage-results} that 
supporting event types other than click 
did not necessarily help the approach 
in reaching significantly more \js code coverage.
This might be explained by the fact that there are more false positives 
in the prediction models for event types other than click,
i.e., the crawler might waste time firing actions on elements 
that do not actually have any events.
Nevertheless, it turns out that supporting these event types
can lead to achieving more code coverage
in shorter time and with fewer crawl actions in certain applications
(e.g., in Chess).

Within 10 minutes time limit, the maximum improvement in \js code coverage 
that \toolName can achieve over the \textit{DEF} strategy 
is 7.91\% (in TacirFormBuilder).
Compared to the \textit{RND} strategy,
the maximum improvement is 10.26\% (in Chess).
With click event only, 
\toolName outperforms the \textit{DEF} strategy by at most 7.91\% (in TacirFormBuilder)
and improves the \textit{RND} strategy by at most 10.22\% (in Chess).

Within 100 crawl actions,
the maximum improvement in \js code coverage that our approach 
can reach is by 8.16\%
compared to the \textit{DEF} strategy (in Calendar),
and by 14.51\% when compared to the \textit{RND} strategy (in Chess).
When considering the click event only,
the maximum improvement
shows up in Calendar:
\toolName can improve \js code coverage
by 17.02\% (compared to \textit{DEF}) and 23.19\% (compared to \textit{RND})
in Google Calendar.

Observe in \Cref{fig:coverage-results} that
\toolName achieves a much higher code coverage in the beginning of the crawling session
in Chess,
while the final code coverage value for both strategies might be close.
We noticed that in Chess the crawler with the \textit{DEF} strategy
keeps performing the same action across different states
by repeatedly clicking on one particular element,
while our approach can overcome this issue by ranking actionables.

In TacirFormBuilder, compared to the \textit{RND} strategy,
\toolName improves the \js code coverage marginally
when the number of actions is limited,
and underperforms when the time is limited.
This is because in this application,
on average, 46.31\% of the DOM nodes are actionable,
and the DOM is considerably small (\Cref{table:experimental-subjects}).
As a result, there is a high chance for a random strategy 
to achieve a high code coverage even in a short time.


\section{Discussion}
\label{sec:discussion}

\header{Limitations}
In this work, we considered only the top-five frequent event types in our dataset.
There are, however, several other event types that can be attached to web elements; our technique can be employed to 
train similar models for other event types.

Another limitation is that our technique is not applicable to web apps 
that use Canvas \html elements.
In theory, however,
our approach should be applicable if the web app uses Salable Vector Graphics (SVG)
for actionables,
since SVG also uses \css for styling.

\header{Relation to existing approaches}
The insight of our approach is that
structural and visual stylistic properties of 
elements can provide cues in exploring a web apps' event space, 
without depending on a specific web browser or an instrumentation technique 
for re-writing event listeners.
There exist techniques 
(e.g., \textsc{Artemis}~\cite{artzi2011framework} or \textsc{JSDep}~\cite{Sung:2016:StaticDomDependencyAnalysis})
that analyze \js code
to explore the event space of web apps.
In theory, these techniques may achieve higher \js code coverage
since they have direct access to the event model of the web app.
However, as we discussed before, it is not always technically possible
or desirable
to use such techniques since they require specialized runtime environments (e.g., browser engines that are modified).
In addition, our approach is not designed to \textit{replace}
such techniques.
Instead, we believe that stylistic features can be helpful in devising, 
for example, hybrid event prioritization techniques. 

We also believe that our work can enhance existing techniques
that focus on state exploration strategies, such as \textsc{FeedEx}~\cite{MilaniFard:2013:FeedEx}, by helping them decide a priori which elements might be actionable.

\header{Threats to validity}
An internal thread to the validity of our evaluation 
is that
executing the crawler on a subject might affect the next crawling sessions.
This is due to the side-effects involving state storage (e.g., cookies, HTML5 local storage),
which can change, for instance, the initial state for the next crawling session.
In such cases, there might be large deviations 
in the code coverage results across different runs of the crawler.
We carefully investigated each app's state after every crawl 
and reverted back any changes made during the previous crawl,
to mitigate this threat.

We included a real-world and complex web app for our evaluation, 
namely Google Calendar,
which is representative for highly-dynamic modern web apps.
A threat in using a proprietary web app in our study is that
its user interface design is likely to change in the future in unknown ways.
This might make replicating the evaluations on this specific subject difficult.
To mitigate this risk, 
we have also included three open-source \js web apps as our subject systems.
These web apps were used in previous studies
related to automated web app exploration too~\cite{artzi2011framework, MilaniFard:2013:FeedEx}.
Moreover, we have provided~\cite{experimental-data} the complete state-flow graph
constructed by the crawler
for all subject systems,
which includes snapshots of the discovered DOM states
in addition to the screenshots of each state,
and the list of actionables and the events that the crawler examined during the crawl
for each of the executions of the crawler.

We included only four subject systems in our evaluation
on \js code coverage.
We believe that these systems can represent both the open source and proprietary web apps.
Also, beside the source code of the open-source subject systems
and the data collected during the execution of the crawler on all subject systems,
we have made available~\cite{experimental-data} the source code of \toolName,
and
the data and R scripts used for training and testing the actionable prediction models,
allowing for replicating the evaluations with more subjects.


\section{Related Work}
\label{sec:related-work}

Many techniques have been proposed 
for crawling dynamic web apps~\cite{Mirtaheri:2013:HistoryOfCrawlers}.
\textsc{AJAXSearch} is tailored for indexing Ajax applications
for efficient searching~\cite{Frey:2007-IndexingAjaxApps, Duda:2008:AjaxSearchTool}.
It only considers \html attributes (e.g., \texttt{onclick}) 
for detecting attached events. 
In contrast, \toolName uses a machine learning model to detect any actionable element regardless of how the events listeners are attached.
\textsc{RE-RIA}~\cite{Amalfitano:2008:ReverseEngineeringRIA} 
collects execution traces from user sessions by manual driving of the web app,
or through automated depth-first crawling in \textsc{CrawlRIA}~\cite{Amalfitano:2010:CrawlRIA}.
It constructs a model from these traces. 
On the contrary, our approach does not need user sessions to drive the web app. 

Peng et al.~\cite{Peng:2012:GraphBased} propose to use a greedy crawling strategy.
The proposed approach assumes that 
the events that change the state of the web application by adding or removing DOM nodes are not fired,
which is far from realistic in modern web apps.
Our technique focuses on exploring the event space of the web application,
rather than providing a novel state exploration strategy.

Dincturk et al.~\cite{Dincturk:2014:ModelBasedForRIA:TWEB} propose a model-based strategy for efficient crawling,
i.e., ``to explore more states early on in the crawling'',
based on an \textit{anticipated model} of the web application.
This model allows to choose events that might lead to more efficient crawling.
The anticipated model adapts to the actual behavior of the web app during the exploration.
A model that reflects the web app's actual behavior as accurately as possible should be chosen,
e.g., a \textit{hypercube}-like model,
which assumes that the newly-explored states share the same events with the old ones.
A crawling strategy with this hypercube-based model is proposed~\cite{Benjamin:2007:StrategyHypercube:Thesis, Benjamin:2011:StrategyHypercube:ICWE};
yet the authors show that the mentioned assumption is frequently violated in real-word apps~\cite{Dincturk:2014:ModelBasedForRIA:TWEB}.
Other proposed meta-models, e.g., the \textit{Menu model}~\cite{Choudhary:2012:MCrawler:Thesis, Choudhary:2013:MenuModel:ICWE},
or the \textit{probabilistic model}~\cite{Dincturk:2012:StatisticalStrategy, Dincturk:2013:ModelBasedCrawlingStrategy} make different assumptions about the web apps under exploration.
In contrast to the model-based crawling, \toolName does not need the user to provide a model of the web application for exploration.
Plus, the implementations of the model-based crawlers require special environments to have full control on \js execution~\cite{Dincturk:2014:ModelBasedForRIA:TWEB}.
\toolName is free from such limitations and can explore any web app, in any web browser, served from any web server.

Moosavi et al.~\cite{Moosavi:2013:ComponentBasedCrawling, Moosavi:2014:ComponentBasedCrawlingICWE} propose the idea of categorizing events for effective state exploration.
However, similar to other approaches, it only considers events attached
via \html attributes.
The approach incorporates \js event handler function names to identify similar events.
\toolName predicts five types of event handlers regardless of how they are attached,
and ranks actionable elements using stylistic cues.
The proposed approach by Moosavi et al. also needs special environments (e.g., browser engines) to run. 

\crawljax~\cite{Mesbah:2012:Crawljax}
is the most cited crawler for modern web apps.
Our technique extends \crawljax's event exploration strategy using stylistic cues.

\textsc{FeedEx}~\cite{MilaniFard:2013:FeedEx}
replaces the default state exploration strategy of \crawljax. 
The strategy sorts partially-explored states using their
\js code coverage, their event path diversity, and DOM diversity.
\textsc{FeedEx} also ranks click events
based on the diversity of the explored states
observed through previous examinations of the events,
i.e., it requires multiple examinations of the same clickable.
In contrast, \toolName focuses on predicting and ranking actionables,
and not on state space exploration strategy, which is the main goal of \textsc{FeedEx}.

\textsc{Artemis}~\cite{artzi2011framework} is a general-purpose
framework for building analysis techniques for dynamic \js web apps.
In contrast to \toolName, \textsc{Artemis} 
requires a special \js engine to execute.
\textsc{JSDep}~\cite{Sung:2016:StaticDomDependencyAnalysis}
improves \textsc{Artemis}
by proposing a technique to identify DOM and event handler dependencies.
Dallmeier et al.~\cite{dallmeier2014webmate, Dallmeier:2013:WebMate} propose a black-box
approach, called \textsc{WebMate}, 
for generating tests for web apps, focusing
on testing cross-browser compatibility issues.
\textsc{WebMate} only considers \js event listeners 
from HTML attributes and for specific libraries (e.g., \textsc{jQuery}).
In contrast to \textsc{WebMate} and \textsc{Artemis}, \toolName is framework- and browser-independent
and focuses on identifying event listeners only using stylistic cues.

Thummalapenta et al.~\cite{sinha:icse2013} propose a method
for guided test generation of web apps
by exercising ``interesting''
behaviors.
An interesting behavior is defined using business rules
and business logic,
which the authors show can be more effective
in covering business rules compared to an undirected technique.
Lin et al.~\cite{Lin:2017:Similarity} propose a technique for exploring the input space
of the web app trough semantic similarity.
We believe that our technique can complement these approaches.

Borges et al.~\cite{Borges:2018:GuidingAppTestingMinedInteractionModels}
propose a technique to learn from interaction models 
to guide mobile app testing.
The proposed approach is similar to ours
in using a machine learning model
to identify interesting elements to click on; 
however, our approach uses a much richer set of stylistic and structural features, 
and also provides a mechanism for
ranking actionables 
in web apps.
Nevertheless,
our approach is also applicable to hybrid mobile apps and 
 mobile (progressive) web apps, which use \css for defining presentation semantics.


\section{Conclusion}
\label{sec:conclusion}

We showed that structural and visual stylistic cues
can aid exploring web apps' event space through crawling.
In particular, our machine learning models
trained on around 700,000 web elements from 1,000 real-world websites
can achieve a high accuracy
in identifying actionables:
90.14\% precision and 87.76\% recall 
when considering the click event listener,
and on average, 
75.42\% precision and 77.76\% recall 
when considering
the five most-frequent event types.
We also devised a technique for ranking actionables
in a crawling session based on stylistic cues.
We concluded that our technique can improve
the \js code coverage achieved by a general-purpose crawler that is tailored for highly-dynamic web apps.

One important direction of future work is to investigate the possibility of 
training machine learning models
for other event types
which are widely used in certain web apps,
e.g., drag and drop.
In addition, we believe that stylistic cues
should be taken into account for
improving other aspects of crawling,
e.g., 
for devising state abstraction functions
that incorporate changes in \css styles.

\bibliographystyle{ACM-Reference-Format}
\bibliography{bibliography}
\balance

\end{document}